# The ubiquity of micron-sized dust grains in the dense interstellar medium

Laurent Pagani[1*], Jürgen Steinacker[1,2], Aurore Bacmann[3], Amelia Stutz[2,4], Thomas Henning[2]

[1]LERMA & UMR 8112 du CNRS, Observatoire de Paris, 61 Av. de l'Observatoire, 75014 Paris, France

[2]Max-Planck-Institut für Astronomie, Königstuhl 17, D-69117 Heidelberg, Germany

[3]Université Joseph Fourier - Grenoble 1 / CNRS, Laboratoire d'Astrophysique de Grenoble (LAOG) UMR 5571, BP 53, 38041 Grenoble Cedex 09, France

[4]Department of Astronomy and Steward Observatory, University of Arizona, 933 North Cherry Avenue, Tucson, AZ 85721, USA

*To whom correspondence should be addressed. E-mail: laurent.pagani@obspm.fr

**Cold molecular clouds are the birthplaces of stars and planets, where dense cores of gas collapse to form protostars. The dust mixed in these clouds is thought to be made of grains of average size 0.1 µm. We report the wide-spread detection of the coreshine effect as a direct sign of the existence of grown, micron-sized dust grains. This effect is seen in half of the cores we have analyzed in our survey, spanning all Galactic longitudes, and is dominated by changes in the internal properties and local environment of the cores, implying that the coreshine effect can be used to constrain fundamental core properties such as the 3D density structure and ages and also the grain characteristics themselves.**

The formation of stars and planets begins in the cold and condensed interstellar medium (*1*) through mechanisms that are not yet well understood (*2*). Thus, much attention is currently paid to characterizing the physical conditions of the dense cloud cores where stars form, before and during their collapse (*3,4,5*) up to the time when a circumstellar disk has formed as the host of future planets (*6*). The materials out of which earth-like planets form are small dust grains embedded in interstellar gas; it has been shown that such grains settle to the disk midplane and grow by coagulation (*7*). In the earlier core phase, coagulation models propose that the dust grains can grow to about micron-size on time scales which depend on the core properties (*8,9*). Furthermore, extinction and emission measurements of cores have been interpreted to contain evidence for the presence of larger grains (*10,11,12,13*). In these models, the difficulty is to disentangle the mixed information about density, dust properties, and temperature.

A recent analysis of the nearby cloud L183 using Spitzer IRAC (*14*) maps revealed unexpected emission at 3.6 μm that is spatially coincident with the densest regions of the cloud (*15*). This pattern can only be accounted for if mid-infrared interstellar radiation is scattered by dust grains in the dense cores larger than the standard grains decribed in (*15,16*). By analogy to the scattered light that has been observed in the outer parts of the cloud at shorter wavelengths and that was named "cloudshine" (*17*), the mid-infrared scattered light from the dense cores has been named "coreshine" (*15*). The maximal grain size visible in the coreshine of L183 is about 1 μm, and could even be larger in the optically thick centre.

We have investigated a sample of 110 cores from the main Spitzer core surveys without any further selection criterion. This survey includes 12 objects with FIR shadow for which a recent study noticed a 3.6 μm excess emission in 7 to 10 of the studied cores (5). From this sample, 95 cores are detectable in the IRAC bands, 50 show clear coreshine, and 8 more cores need further analysis to confirm possible coreshine. Thus, about half of all low-mass cores in our sample show this effect. The fluxes range from 0.01 up to 0.5 MJy/sr, compared to the background field peak flux at 3.5 μm of about 12 MJy/sr. Some of the most prominent examples of sources with coreshine are displayed in Fig. 1. The dust in the core absorbs the background radiation at 8 μm (right image for each core) while it also scatters the external radiation field at 3.6 μm (left image for each core). The good spatial correlation of the scattered light flux at 3.6 μm with the extinction at 8.0 μm shows that the coreshine comes from the dense parts of the molecular cloud cores.. These examples include cores with simple geometry like L1507, binary cores like L1262, filamentary structures like L1251A, cores with infall signature like L1544 (*18*), with a proto-stellar object like L1521F, or cores with jet activity like L1157. Thus, coreshine can appear with or without the presence of a Class 0, Class I protostar or of a strong outflow.

Aside from scattered light, the emission at mid-infrared wavelengths could also come from stochastically heated small dust grains like polycyclic aromatic hydrocarbons (PAHs) (*19*). However, an excitation of these grains requires exposure to the UV and optical interstellar radiation field (ISRF). This radiation is absorbed and scattered in the outer edges of the clouds, while the coreshine comes from the inner dense regions lacking any UV or optical illumination (*15*). The ρ Oph 9 region (*20*) is an example of a cloud with two cores where both emission types can be seen in the four IRAC bands observed with the Spitzer telescope (Fig. 2). While the PAH emission of the bright rim to the left is visible especially at 5.8 and 8 μm, the 3.6 μm coreshine around the small core to the lower right comes from the extincted region mapped at 8 μm. The flux ratio 8/3.6 μm (Fig. 2) reveals the well-correlated coreshine and extinction region.

While the coreshine effect is common among different types of cores, it is also present across a variety of star-forming environments (Fig. 3). Most nearby star-forming regions contain cores with and without coreshine which indicates that the appearance of larger grains depends on the local properties and history of the core. Only in the Gum/Vela nebula region, no coreshine is detected so far. It is possible that a fraction of the large grains in this region might have been destroyed by a supernova that is thought to have reshaped the region about 1 Million years ago (*21*). There are not many cases of local cores with detectable coreshine close to the Galactic plane (Fig.3). It is not clear presently whether this is due to the increased number of background stars confusing the coreshine signal or physical effects which cause changes in the dust opacities with Galactic position (*22*).

To investigate the influence of the anisotropic illumination on the coreshine appearance with varying Galactic position, we performed 3D radiative transfer calculations (*23*) for a 10 solar-mass core (Fig. 4). We used a radial density power-law profile in the outer parts (index -1.8) and a flattening towards the centre, in agreement with core density profile estimates (*3*). Two different cases were considered, either with a flattening in the innermost 4000 A.U., or with a more centrally concentrated density profile, flattening only in the innermost 1000 A.U. The coreshine flux in the core with a large flattening varies little with Galactic core position. However, our model shows a slight increase in flux in the direction towards the Galactic Center in the four cases where the core is viewed at 90 deg from the Galactic Center line-of-sight. More prominently, the case with a more concentrated density profile reproduces the central flux depression that is seen in a few dense cores like L183 and L1544, where the optical depth at 3.6 μm becomes too large for scattered photons to escape. In the investigated sample, only a handful of cores show this inner flux depression. The asymmetrical flux increase for the 4 cores 90° from the Galactic Center line of sight is also more pronounced than in the flatter profile cases. In both types of profile, for the central image where the core is placed in front of the Galactic center, the photons are scattered in forward direction but have to cross the core so that the flux is smaller than for the side illumination cases. In general, the coreshine effect is therefore only weakly dependent upon the location of the source in the Galaxy and is dominated by the properties of the core.

The coreshine effect thus provides a direct tool to investigate the grain and ice-mantle growth process for which turbulence and density are major ingredients (*9*). Because the growth process is continuous, it can serve as a measure of the age of cores, contrary to chemical "clocks" which may suffer from repetitive resets (*1*). A "grain growth clock" may help distinguishing between the currently proposed models (*1*) to stabilize pre-stellar cores before they collapse.

In the optically thin case, the coreshine flux has the potential to independently provide a measurement of the dust column density. Because scattering is highly sensitive to grain size (*15*), the coreshine effect will help improve our understanding of the grain infra-red opacities. Moreover, the combined consideration of scattering, extinction, and emission modeling will constrain the detailed 3D structure of cold molecular clouds down to the cores.

The growth of grains has an impact on the ion-neutral cold chemistry within the core via electronic equilibrium and possibly surface reactions, while circumstellar disks might be affected by a seed population already containing large particles as revealed by the coreshine effect. More generally, the grains unveiled by coreshine might help to characterize the recent history of an entire region, as in the case of the Gum Nebula complex, where an absence of coreshine implies recent destruction of big dust grains, possibly by a supernova explosion.

25. This work is based in parts on observations made with the Spitzer Space Telescope, which is operated by the Jet Propulsion Laboratory, California Institute of Technology under a contract with NASA. J.S. is grateful for the hospitality and financial support of the Observatoire de Paris and LERMA, where some of this work was performed. We also thank two anonymous reviewers. The first two authors have had an equal share of the part of the work they did.

**images**

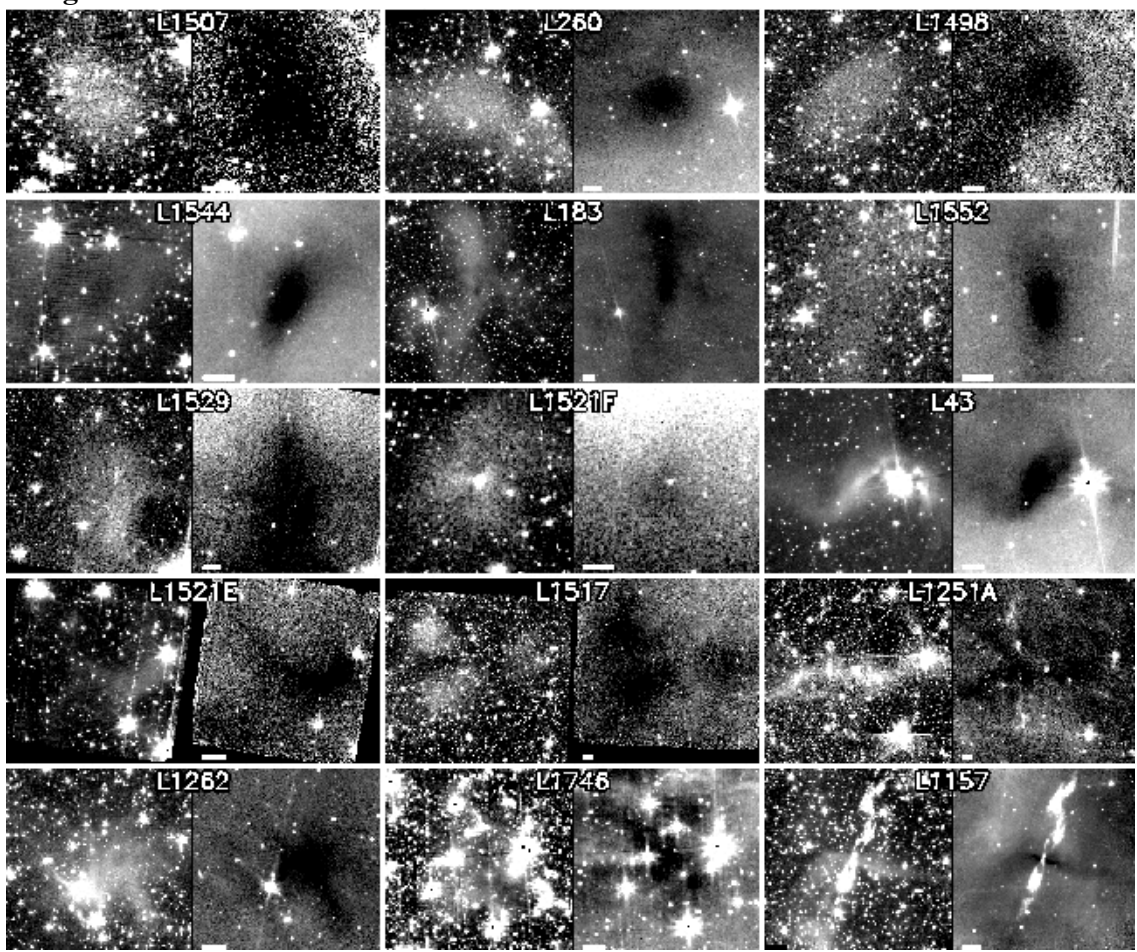

**Fig.1 15 cores with prominent coreshine**. For each core, the IRAC 3.6 (left) and 8.0 μm (right) Spitzer maps are shown. The white bar in the bottom left corner of the 8 μm images represents a length of 4,000 AU (1 AU = $1.496 \times 10^{11}$ m).

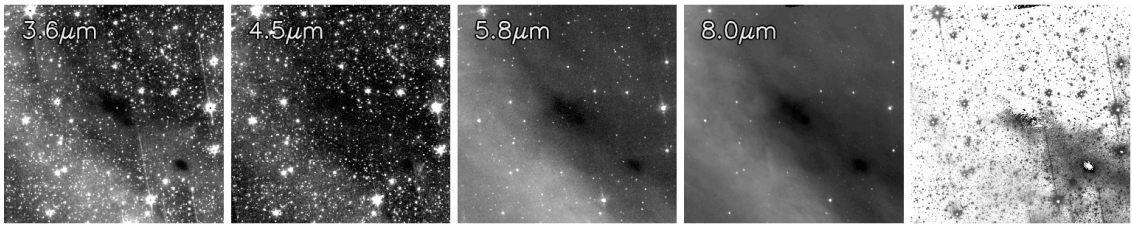

**Fig.2 Coreshine and PAH emission in ρ Oph 9**. Left: The four IRAC Spitzer images at 3.6, 4.5, 5.8, and 8.0 µm. Right: The flux ratio of the images at 8 and 3.6 µm reveals the well-correlated coreshine and extinction region. The size of each frame is 128,000 AU.

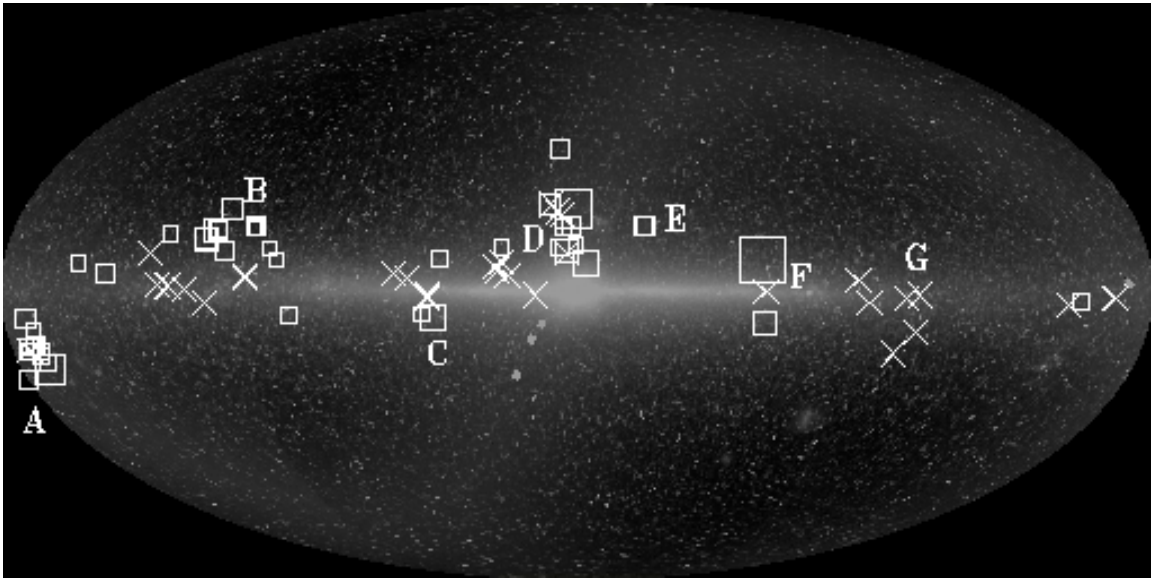

**Fig.3 Spatial distribution of cores with and without the coreshine effect**. The positions of the investigated cores are plotted over the interstellar radiation allsky map at 3.5 µm (DIRBE, *24*), representing the background field which is scattered by grains. The centre of the Galaxy is at the centre of the map, the anti-centre is at the left and right borders. The coreshine flux is indicated as squares scaling logarithmically with the flux. Crosses indicate cores without measurable coreshine flux. Well-known star formation regions are annotated by letters: A=Taurus, B=Cepheus, C=Aquila, D=Ophiuchus, E=Lupus, F=Crux, G=Vela.

939

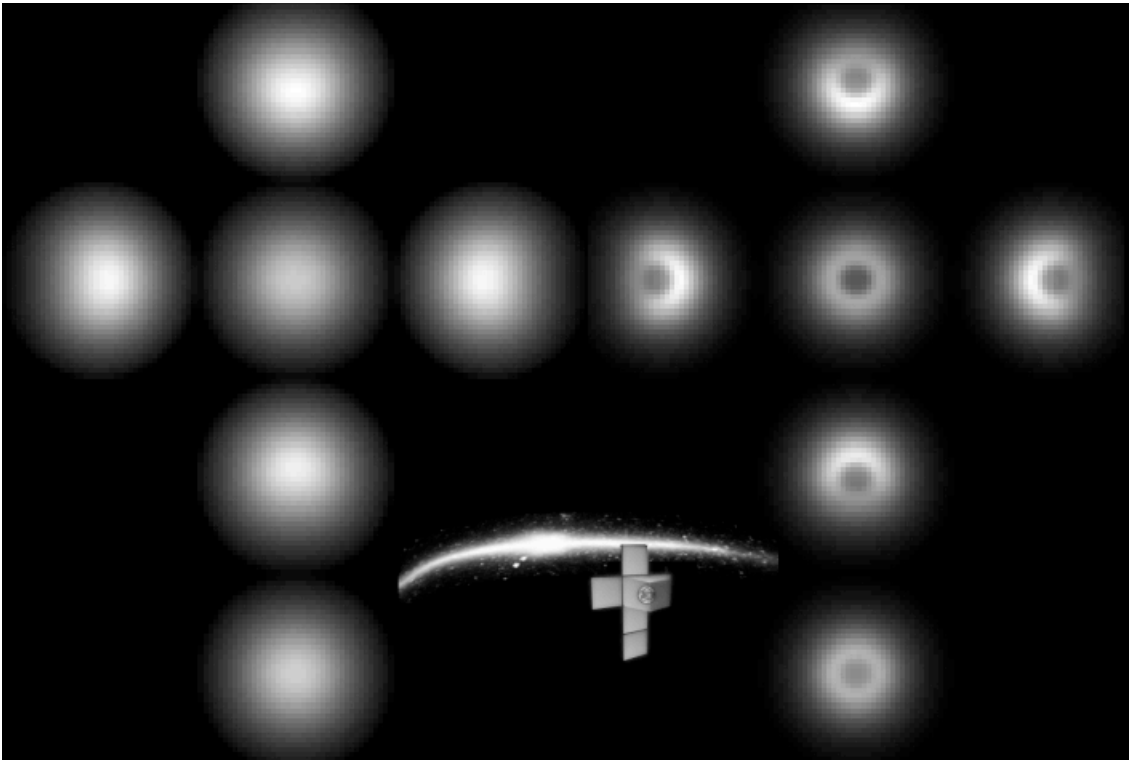

**Fig.4 Coreshine model across the Galaxy**. Scattered light radiative transfer modeling images of the coreshine pattern of a molecular cloud core. The core is placed on a rectilinear coordinate system where one axis is the line from the Earth to the Galactic Centre (therefore the lowest image is the view to the galactic anti-centre). Coreshine for a 10 solar mass core of diameter 20,000 AU. Left: with a profile flattening at a radius of 4,000 AU. Right: with a flattening at 1,000 AU.